\documentclass[onecolumn,amsmath,amssymb,showpacs,preprint]{revtex4}
\usepackage{mathtext,bm,bbm,amsmath,amsfonts,amssymb,indentfirst,syntonly,graphicx}
\usepackage{graphicx}
\usepackage{epsfig}
\usepackage{graphicx}

\begin{document}

\title{Testing the anisotropy of the universe using the simulated gravitational wave events from advanced LIGO and Virgo}

\author{Hai-Nan Lin $^1$}
\email{linhn@ihep.ac.cn}
\author{Jin Li $^{1}$}
\email{cqujinli1983@cqu.edu.cn}
\author{Xin Li $^{1}$}
\email{lixin1981@cqu.edu.cn}
\affiliation{$^1$Department of Physics, Chongqing University, Chongqing 401331, China}

\begin{abstract}
  The detection of gravitational waves (GWs) provides a powerful tool to constrain the cosmological parameters. In this paper, we investigate the possibility of using GWs as standard sirens in testing the anisotropy of the universe. We consider the GW signals produced by the coalescence of binary black hole systems and simulate hundreds of GW events from the advanced Laser Interferometer Gravitational-Wave Observatory (LIGO) and Virgo. It is found that the anisotropy of the universe can be tightly constrained if the redshift of the GW source is precisely known. The anisotropic amplitude can be constrained with an accuracy comparable to the Union2.1 complication of type-Ia supernovae if $\gtrsim 400$ GW events are observed. As for the preferred direction, $\gtrsim 800$ GW events are needed in order to achieve the accuracy of Union2.1. With 800 GW events, the probability of pseudo anisotropic signals with an amplitude comparable to Union2.1 is negligible. These results show that GWs can provide a complementary tool to supernovae in testing the anisotropy of the universe.
\end{abstract}

\maketitle

\section{Introduction}

The cosmological principle, which states that the universe is homogeneous and isotropic on large scales, is one of the most basic assumptions of modern cosmology. This assumption is proven to be well consistent with various observations, such as the statistics of galaxies \cite{TrujilloGomez:2010yh}, the halo power spectrum \cite{Reid:2009xm}, the observation on the growth function \cite{Nesseris:2007pa}, the cosmic microwave background from the Wilkinson Microwave Anisotropy Probe (WMAP) \cite{Bennett:2011,Bennett:2013} and Planck satellites \cite{Ade:2013zuv,Ade:2015xua}. Based on the cosmological principle, the standard cosmological model, i.e. the cold dark matter plus a cosmological constant ($\Lambda$CDM) model is well constructed. However, some other observations show that the universe may deviate from the statistical anisotropy. These include but not limited to the large scale bulk flow \cite{Kashlinsky:2008ut,Lavaux:2008th}, the CMB temperature anisotropy \cite{Ade:2014nvu,Ade:2016lfjf}, the spatial variation of the electromagnetic fine-structure constant \cite{Murphy:2003hw,Webb:2010hc,King:2012id,Pinho:2016mkm}, the anisotropy of the distance-redshift relation of type-Ia supernovae \cite{Antoniou:2010gw,Mariano:2012wx,Kalus:2012zu}. If the universe is indeed anisotropic, it implies that there are new physics beyond the standard model. Whether these anisotropic signals come from the intrinsic property of the universe or merely the statistical fluctuation is extensively debated\cite{Jimenez:2014jma,Quartin:2014yaa,Appleby:2014lra,Whitmore:2014ina,Chang:2015jhd,Lin:2016nlm}.

The gravitational waves (GWs) provide an alternative tool to testing the cosmology. The greatest advantages of GWs is that the distance calibration is independent of any other distance ladders, i.e. it is self-calibrating. Since Einstein predicted the existence of GWs a century ago, extensive efforts have been made to directly detect GWs but without success. The breakthrough happens in September 2015, when the Laser Interferometer Gravitational-Wave Observatory (LIGO) and Virgo collaborations reported a GW signal produced by the coalescence of two black holes, which was late named GW150914 \cite{Abbott:2016a}. Since then, four more GW events have been observed \cite{Abbott:2016b,Abbott:2017a,Abbott:2017b,Abbott:2017c}. The first four events are produced by the merge of binary black hole systems and no electromagnetic counterpart is expected. The last one event, GW170817, is produced by the merge of binary neutron star system and it is associated with a short gamma-ray burst GRB170817 \cite{Abbott:2017d,Goldstein:2017,Savchenko:2007}. The host galaxy NGC4993 at redshift $z\sim 0.01$ is identified by the follow-up observation \cite{Coulter:2017wya}. The simultaneous observations of GW signal and electromagnetic counterparts open the new era of multi-messenger astronomy. Using the GW/GRB170817 event as standard siren, the Hubble constant is constrained to be $70.0_{-8.0}^{+12.0}~{\rm km~s}^{-1}~{\rm Mpc}^{-1}$ \cite{Abbott:2017xzu}, showing that GW data are very promising in constraining the cosmological parameters. Several works have used the simulated GW data to constrain the cosmological parameters and showed that the constraint ability of GWs is comparable or even better than the traditional probes if hundreds of GW events have been observed \cite{Sathyaprakash:2009xt,DelPozzo:2011yh,Zhao:2010sz,DelPozzo:2017bna,Cai:2017sby,Cai:2017aea}.

In this paper, we investigate the possibility of using GW data to test the anisotropy of the universe. Unfortunately, there is only five GW events observed up to date. With such a small amount of data points, it is impossible to do statistical analysis. Therefore, we simulate a large number of GW events from the advanced LIGO and Virgo detectors. It is expected that hundreds of GW events will be detected in the next years. We use the simulated GW data to test how many GW events are needed in order to reach the accuracy of type-Ia supernovae. The present astronomical observations imply that the intrinsic anisotropy of the universe is quite small and could be treat as a perturbation of $\Lambda$CDM model. Therefore, throughout this paper we assume a fiducial flat $\Lambda$CDM model with Planck parameters $\Omega_M=0.308$ and $H_0=67.8~{\rm km~s}^{-1}~{\rm Mpc}^{-1}$\cite{Ade:2015xua}.

The rest of the paper is organized as follows: In section \ref{sec:methodology}, we describe the method of using GW data as standard sirens in cosmological studies. In section \ref{sec:simulation}, we illustrate how to simulate the GW events from the advanced LIGO and Virgo. In section \ref{sec:results}, we investigate the constraint ability of GW data on the anisotropy of the universe. Finally, discussions and conclusions are given in section \ref{sec:conclusions}.

\section{GWs as standard sirens}\label{sec:methodology}

GW is the fluctuation of spacetime metric, as a prediction of general relativity it has two polarization states  often written as $h_+(t)$ and $h_\times(t)$. GW detectors based on the interferometers such as advanced LIGO and Virgo measure the change of difference of two optical path (which is often called the strain) caused by the pass of GWs. The strain is the linear combination of the two polarization states,
\begin{equation}
  h(t)=F_+(t)h_+(t)+F_\times(t)h_\times(t),
\end{equation}
where the coefficients $F_+(t)$ and $F_\times(t)$ are called the beam-pattern functions, which depends on the location and orientation of the detector, as well as the position of GW source. For detectors built on the Earth, due to the diurnal motion of the Earth the beam patterns are periodic functions of time with a period equal to one sidereal day. The explicit expressions of the beam patterns are given by \cite{Jaranowski:1998qm}
\begin{eqnarray}
  F_+(t)=\sin\zeta[a(t)\cos2\psi+b(t)\sin2\psi],\\
  F_\times(t)=\sin\zeta[b(t)\cos2\psi-a(t)\sin2\psi],
\end{eqnarray}
where $\zeta$ is the angle between the two interferometer arms, $\psi$ is the polarization angle of GW, and
\begin{eqnarray}\nonumber
  a(t)&=&\frac{1}{16}\sin2\gamma(3-\cos2\lambda)(3-\cos2\delta)\cos[2(\alpha-\phi_r-\Omega_rt)]\\ \nonumber
  &&-\frac{1}{4}\cos2\gamma\sin\lambda(3-\cos2\delta)\sin[2(\alpha-\phi_r-\Omega_rt)]\\ \nonumber
  &&+\frac{1}{4}\sin2\gamma\sin2\lambda\sin2\delta\cos[\alpha-\phi_r-\Omega_rt]\\ \nonumber
  &&-\frac{1}{2}\cos2\gamma\cos\lambda\sin2\delta\sin[\alpha-\phi_r-\Omega_rt]\\
  &&+\frac{3}{4}\sin2\gamma\cos^2\lambda\cos^2\delta,
\end{eqnarray}
\begin{eqnarray}\nonumber
  b(t)&=&\cos2\gamma\sin\lambda\sin\delta\cos[2(\alpha-\phi_r-\Omega_rt)]\\ \nonumber
  &&+\frac{1}{4}\sin2\gamma(3-\cos2\lambda)\sin\delta\sin[2(\alpha-\phi_r-\Omega_rt)]\\ \nonumber
  &&+\cos2\gamma\cos\lambda\cos\delta\cos[\alpha-\phi_r-\Omega_rt]\\
  &&+\frac{1}{2}\sin2\gamma\sin2\lambda\cos\delta\sin[\alpha-\phi_r-\Omega_rt],
\end{eqnarray}
where $\gamma$ is measured counterclockwise from East to the bisector of the interferometer arms, $\lambda$ is the latitude of the detector's location, $(\alpha,\delta)$ are the right ascension and declination of the GW source in the equatorial coordinate system, $\Omega_r$ is the rotational angular velocity of the Earth, and $\phi_r$ is the initial phase characterizing the position of the Earth in its diurnal motion at $t=0$. Therefore, $\phi_r+\Omega_rt$ represents the local sidereal time of the detector's location. For GW transients such as the five events observed by advanced LIGO and Virgo, the duration of GW signal is much smaller than one sidereal day. In such a short time the motion of the Earth can be neglected and the pattern functions are approximately time-independent.

In this paper, we focus on the GW signals produced by the coalescence of binary systems. Consider a binary system consists of component masses $m_1$ and $m_2$ in the comoving frame, define the total mass $M=m_1+m_2$, the symmetric mass ratio $\eta=m_1m_2/M^2$, and the chirp mass $\mathcal{M}_c=M\eta^{3/5}$. For the GW source locating at cosmological distance with redshift $z$, the chirp mass in the observer frame is given by $\mathcal{M}_{c,{\rm obs}}=(1+z)\mathcal{M}_{c,{\rm com}}$ \cite{Krolak:1987}. In the following, $\mathcal{M}_c$ always refers to the chirp mass in the observer frame unless otherwise stated. In the post-Newtonian and stationary phase approximation, the strain $h(t)$ produced by the inspiral of binary, is given in the Fourier space by \cite{Sathyaprakash:2009,Zhao:2010sz}
\begin{equation}\label{eq:fourier_strain}
  \mathcal{H}(f)=\mathcal{A}f^{-7/6}\exp[i(2\pi f t_0-\pi/4+2\psi(f/2)-\varphi_{(2,0)})],
\end{equation}
where $t_0$ is the epoch of merger. The explicit expressions of the phase terms $\psi(f/2)$ and $\varphi_{(2,0)}$ can be found in Ref.\cite{Sathyaprakash:2009}, but these are unimportant in the following calculation because we are only interested in the inner product of $\mathcal{H}(f)$, so the exponential term on the right-hand-side of equation (\ref{eq:fourier_strain}) is canceled out. The Fourier amplitude $\mathcal{A}$ is given by
\begin{equation}
  \mathcal{A}=\frac{1}{d_L}\sqrt{F_+^2(1+\cos^2\iota)^2+4F_\times^2\cos^2\iota}\times\sqrt{\frac{5\pi}{96}}\pi^{-7/6}\mathcal{M}_c^{5/6},
\end{equation}
where $\iota$ is the inclination of the binary's orbital plane, i.e. the angle between the binary's orbital angular momentum and the line-of-sight, and
\begin{equation}\label{eq:dL}
  d_L=\frac{1+z}{H_0}\int_0^z\frac{dz}{\sqrt{\Omega_M(1+z)^3+1-\Omega_M}}
\end{equation}
is the luminosity distance of the GW source to the detector.

The signal-to-noise ratio (SNR) of a detector is given by the square root of the inner product of the strain in Fourier space \cite{Sathyaprakash:2009},
\begin{equation}\label{eq:snr}
  \rho_i=\sqrt{\langle \mathcal{H},\mathcal{H}\rangle},
\end{equation}
where the inner product is defined as
\begin{equation}
  \langle a,b \rangle=4\int_{f_{\rm lower}}^{f_{\rm upper}}\frac{\tilde{a}(f)\tilde{b}^*(f)+\tilde{a}^*(f)\tilde{b}(f)}{2}\frac{df}{S_h(f)},
\end{equation}
where $\tilde{}$ represents the Fourier transformation and * represents the complex conjugation, $S_h(f)$ is the one-side noise power spectral density (PSD) characterizing the sensitivity of the detector on the spacetime strain, $f_{\rm lower}$ and $f_{\rm upper}$ are the lower and upper cutoffs of the frequency. Bellow $f_{\rm lower}$ the noise is uncontrollable and $S_h(f)$ is often assumed to be infinity. $f_{\rm upper}$ is the highest frequency of the GW signal during the inspiral epoch. Following Ref.\cite{Zhao:2010sz}, we assume $f_{\rm upper}=2f_{\rm LSO}$, where $f_{\rm LSO}=1/(6^{3/2}2\pi M_{\rm obs})$ is the orbit frequency at the last stable orbit, $M_{\rm obs}=(1+z)(m_1+m_2)$ is the total mass in observer frame. If $N$ independent detectors form a network and detect the same GW source simultaneously, the combined SNR is given by
\begin{equation}\label{eq:snr_combined}
  \rho=\left[\sum_{i=1}^N\rho_i^2\right]^{1/2}.
\end{equation}
We require the SNR to be larger than 8 to ensure that this is indeed the GW signal rather than the noise.

The uncertainty of luminosity distance extracted from the GW signals can be obtained using the Fisher matrix \cite{Zhao:2010sz}. Note that the distance of source $d_L$ is correlated with other parameters, especially the inclination angle $\iota$. In principle, all values of $\iota\in [0^{\circ}, 180^{\circ}]$ are possible. It is pointed out that the maximal effect of inclination angle on the SNR is a factor of 2 \cite{Cai:2017sby,Li:2015nhd}. Here we only consider the simplified case where the binary's orbital plane is nearly face on, hence the amplitude $\mathcal{A}$ is independent of the polarization angle $\psi$.  Following Ref.\cite{Cai:2017sby}, we assume that $d_L$ is uncorrelated with that of other parameters and then double the uncertainty of $d_L$ calculated from the Fisher matrix as the upper limit of instrument error on $d_L$, i.e.,
\begin{equation}
  \sigma_{d_L}^{\rm inst}=\frac{2d_L}{\rho}.
\end{equation}
Such a treatment, although is not accurate, is reasonable because we are only interested in the constraining ability of GW events on the anisotropy of the universe. A more accurate measurement of distance gives a tighter constraint on the anisotropy. We also add an additional uncertainty $\sigma_{d_L}^{\rm lens}=0.05zd_L$ caused by the weak lensing of galaxies alone the line-of-sight. Therefore, the total error on $d_L$ is given by
\begin{equation}\label{eq:dL_error}
  \sigma_{d_L}=\sqrt{\left(\frac{2d_L}{\rho}\right)^2+(0.05zd_L)^2}\,.
\end{equation}

The luminosity distance is often converted to the dimensionless distance modulus by
\begin{equation}\label{eq:mu}
  \mu=5\log\frac{d_L}{\rm Mpc}+25,
\end{equation}
and the uncertainty of $\mu$ is propagated from that of $d_L$ by
\begin{equation}\label{eq:mu_error}
  \sigma_\mu=\frac{5}{\ln 10}\frac{\sigma_{d_L}}{d_L}.
\end{equation}
To use GWs as the standard sirens to test anisotropy of the universe, we assume that the universe has dipole structure in distance-redshift relation, i.e.
\begin{equation}\label{eq:dipole_mu}
  \mu=\mu_{\rm \Lambda CDM}(1-d\cos\theta),
\end{equation}
where $d$ is the dipole amplitude and $\theta$ is the angle between GW source and the preferred direction of the universe. The preferred direction can be parameterized as $(\alpha_0,\delta_0)$ in the equatorial coordinate system. The three parameters $(d,\alpha_0,\delta_0)$ can be obtained by fitting the GW data to equation (\ref{eq:dipole_mu}) using the least-$\chi^2$ method.

\section{Simulation from advanced LIGO and Virgo}\label{sec:simulation}

In this section we simulate GW events based on the advanced LIGO and Virgo detectors. The advanced LIGO consists of two detectors locating at Hanford, WA $(119.41^{\circ}~{\rm W},46.45^{\circ}~{\rm N})$ and Livingston, LA $(90.77^{\circ}~{\rm W},30.56^{\circ}~{\rm N})$, respectively. Each detector contains a laser interferometer with two orthogonal arms of about 4 kilometers. The Virgo detector has two arms of three kilometres long and locates near Pisa, Italy $(10.50^{\circ}~{\rm E},43.63^{\circ}~{\rm N})$. The instrument parameters are \cite{Jaranowski:1998qm}: $\lambda_H=46.45^{\circ}$, $\gamma_H=171.80^{\circ}$, $\lambda_L=30.56^{\circ}$, $\gamma_L=243.00^{\circ}$, $\lambda_V=43.63^{\circ}$, $\gamma_V=116.50^{\circ}$ and $\zeta_H=\zeta_L=\zeta_V=90^{\circ}$. The symbols have been clarified in the last section, and the subscripts `H', 'L' and `V' stand for Hanford, Livingston and Virgo respectively. Since Hanford locates at the west of Livingston by longitude difference $28.64^{\circ}$, the local sidereal time of Hanford is later than that of Livingston by $28.64^{\circ}$ (corresponding to 1.91 hours), i.e.
\begin{equation}\label{eq:delta_t1}
  (\phi_r+\Omega_rt)_H-(\phi_r+\Omega_rt)_L=-28.64^{\circ}.
\end{equation}
Similarly, we have
\begin{equation}\label{eq:delta_t2}
  (\phi_r+\Omega_rt)_H-(\phi_r+\Omega_rt)_V=-129.91^{\circ}.
\end{equation}
The PSD of the advanced advanced LIGO is given by\cite{Ajith:2009},
\begin{equation}
  S_h(f)=S_0\left[x^{-4.14}-5x^{-2}+\frac{111(1-x^2+0.5x^4)}{1+0.5x^2}\right],
\end{equation}
where $x=f/f_0$, $f_0=215$ Hz, $S_0=1.0\times 10^{-49}~{\rm Hz}^{-1}$, and $f_{\rm lower}=20$ Hz. The PSD of the advanced Virgo is given by \cite{Ajith:2009}
\begin{equation}
  S_h(f)=S_0\left[2.67\times 10^{-7}x^{-5.6}+0.59x^{-4.1}\exp(-\alpha)+0.68x^{5.34}\exp(-\beta)\right],
\end{equation}
where $x=f/f_0$, $f_0=720$ Hz, $S_0=1.0\times 10^{-47}~{\rm Hz}^{-1}$, $\alpha=(\ln x)^2[3.2+1.08\ln x+0.13(\ln x)^2]$, $\beta=0.73(\ln x)^2$ and $f_{\rm lower}=20$ Hz.

We consider the merge of binary black hole systems as the source of GW. The mass of each black hole is assumed to be uniformly distributed in the range $[3,100]M_{\odot}$. We also require that the mass difference of two component black holes is not to large, and restrict the mass ratio $q=m_1/m_2$ in the range [0.5,2.0]. Furthermore, we assume that the sources are randomly distributed in the sky. Taking the time evolution of the burst rate into consideration, the probability distribution function of GW sources reads \cite{Zhao:2010sz}
\begin{equation}\label{eq:pdf_redshift}
  P(z)\propto \frac{4\pi d_C^2(z)R(z)}{H(z)(1+z)},
\end{equation}
where $H(z)=H_0\sqrt{\Omega_M(1+z)^3+1-\Omega_M}$ is the Hubble parameter, $d_C=\int_0^z1/H(z)dz$ is the comoving distance, $R(z)=1+2z$ for $z\leqslant 1$, $R(z)=(15-3z)/4$ for $1<z<5$, and $R(z)=0$ otherwise. The advanced LIGO and Virgo is expected to be able to detect GW signals produced by the merge of binary black hole at reshift $z\sim 1$ if it reaches to the ultimately designed sensitivity \cite{Sathyaprakash:2009}.

We simulate a set of GW events, each of which contains the parameters $(z,\alpha,\delta,\mu,\sigma_{\mu})$ that we are interested in. Since we will use the simulated data to study the anisotropy of the universe, the sky position of GW source $(\alpha,\delta)$ are required. The detailed procedures of simulation are as follows:
\begin{enumerate}
  \item Sample black hole mass from uniform distribution, $m_1,m_2\in U[3,100]M_\odot$.
  \item If $0.5<m_1/m_2<2.0$, sample redshift $z$ from the probability distribution function (\ref{eq:pdf_redshift}); else go back to step 1.
  \item Sample the sky position ($\alpha,\delta$) from the uniform distribution on 2-dimensional sphere. This can be done by sampling $\alpha\in U[0,2\pi]$, $x\in U[-1,1]$, and setting $\delta=\arcsin x$.
  \item Sample $(\phi_r+\Omega_rt)_H$ from $U(0,2\pi)$, and calculate $(\phi_r+\Omega_rt)_L$ and $(\phi_r+\Omega_rt)_V$ from equations (\ref{eq:delta_t1}) and (\ref{eq:delta_t2}).
  \item Calculate the SNR of each detector $\rho_H$, $\rho_L$ and $\rho_V$ using equation (\ref{eq:snr}), and the combined SNR $\rho$ using equation (\ref{eq:snr_combined}).
  \item If $\rho>8$, calculate the fiducial luminosity distance $d_L$ and its uncertainty $\sigma_{d_L}$ from equations (\ref{eq:dL}) and (\ref{eq:dL_error}), respectively; else go back to step 1.
  \item Convert $d_L$ and $\sigma_{d_L}$ to $\mu$ and $\sigma_\mu$ using equations (\ref{eq:mu}) and (\ref{eq:mu_error}). Calculate the anisotropic distance modulus $\mu$ using equation (\ref{eq:dipole_mu}).
  \item Sample the simulated distance modulus from Gaussian distribution $\mu_{\rm sim}\sim\mathcal{G}(\mu,\sigma_\mu)$. Go back to step 1, until the desired number of GW events are obtained.
\end{enumerate}

Some issues should be clarified. Since we only consider the case where the binary's orbital plane is nearly face on, our result does not depend on the polarization angle $\psi$. The GW events can happen at any time during the run of detectors. We assume that one GW event is uncorrelated with other GW events, and the GW events are uniformly distributed with time. Due to the periodicity of the diurnal motion, we can sample $(\phi_r+\Omega_rt)_H$ from $U(0,2\pi)$, whereas $(\phi_r+\Omega_rt)_L$ and $(\phi_r+\Omega_rt)_V$ are determined by the difference of local sidereal times of detectors' locations. To ensure the significance of GW signals, we require the combined SNR of three detectors to be larger than 8. In step 7, when calculating the anisotropic distance modulus, we choose the fiducial dipole amplitude to be the same to Union2.1, i.e. $d=1.2\times 10^{-3}$ \cite{Lin:2016nlm}. The fiducial direction is arbitrarily chosen to be $(\alpha_0,\delta_0)=(310.6^{\circ},-13.0^{\circ})$, which is also the same to Union2.1 but changed from the galactic coordinates to equatorial coordinates.

\section{Constrain on anisotropy}\label{sec:results}

In this section we use the simulated GW data to constrain the anisotropy of the universe. We want to see, with current accuracy of detectors, how many GW events are needed in order to correctly reproduce the fiducial dipole amplitude and direction. We simulate $N=100,200,300,\cdots,900$ data points, respectively. For each $N$ we repeat the simulation 1000 times and calculate the average dipole amplitude and preferred direction.

\begin{figure}[htbp]
  \centering
  \includegraphics[width=0.6\textwidth]{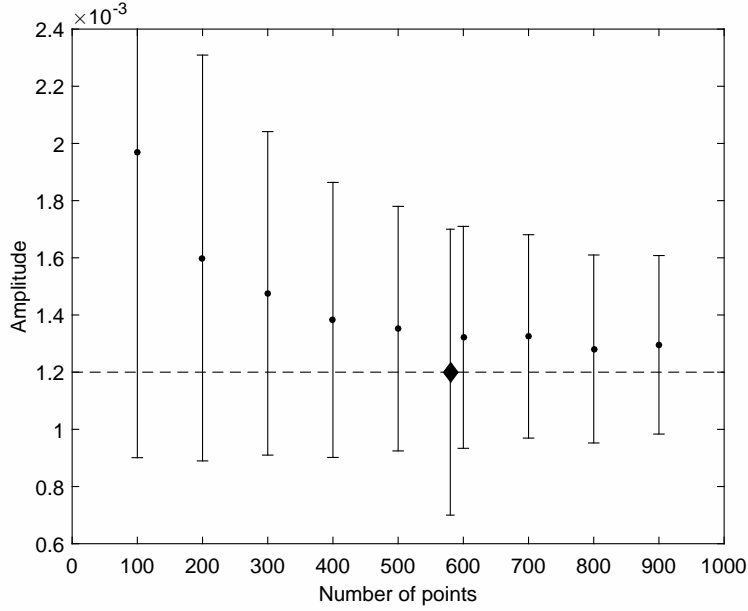}\\
  \caption{The average dipole amplitude and $1\sigma$ error as a function of the number of GW events. The black diamond is the dipole amplitude of Union2.1. The dashed line is the fiducial dipole amplitude.}\label{fig_aniso_all_d}
\end{figure}

In Fig.\ref{fig_aniso_all_d}, we plot the average dipole amplitude as a function of the number of GW events $N$.
The central value is the mean of dipole amplitudes in 1000 simulations, and the error bar is the root mean square of uncertainty of dipole amplitudes in 1000 simulations. The dipole amplitude of Union2.1 is also plotted for comparison. From this figure, we can see that, as the number of GW events increases, the constrained dipole amplitude gets more close to the fiducial dipole amplitude, and at the same time the uncertainty is reduced. To reach the accuracy of Union2.1, $\gtrsim 400$ GW events are needed.

\begin{figure}[htbp]
  \centering
  \includegraphics[width=0.6\textwidth]{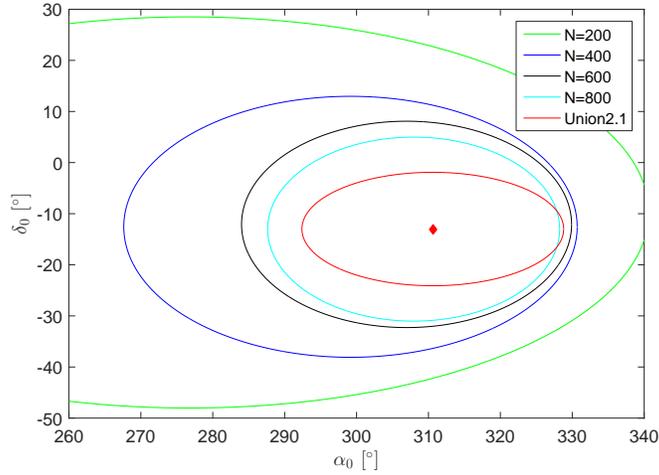}\\
  \caption{The average $1\sigma$ confidence region in the $(\alpha_0,\delta_0)$ plane for different number of GW events. The $1\sigma$ confidence region of Union2.1 is also plotted for comparison.}\label{fig_aniso_all_lb}
\end{figure}

To see the constraint ability of GWs on the preferred direction, in Fig.\ref{fig_aniso_all_lb} we plot the average $1\sigma$ confidence region in the $(\alpha_0,\delta_0)$ plane for different $N$.
For comparison the $1\sigma$ confidence region of Union2.1 is also plotted. This figure shows that $\sim 400$ GW events are not enough to tightly constrain the preferred direction. More than 800 events are needed in order to reach the accuracy of Union2.1. As is expected, increasing the number of GW events can tighten the constraint.

\begin{figure}[htbp]
  \centering
  \includegraphics[width=0.6\textwidth]{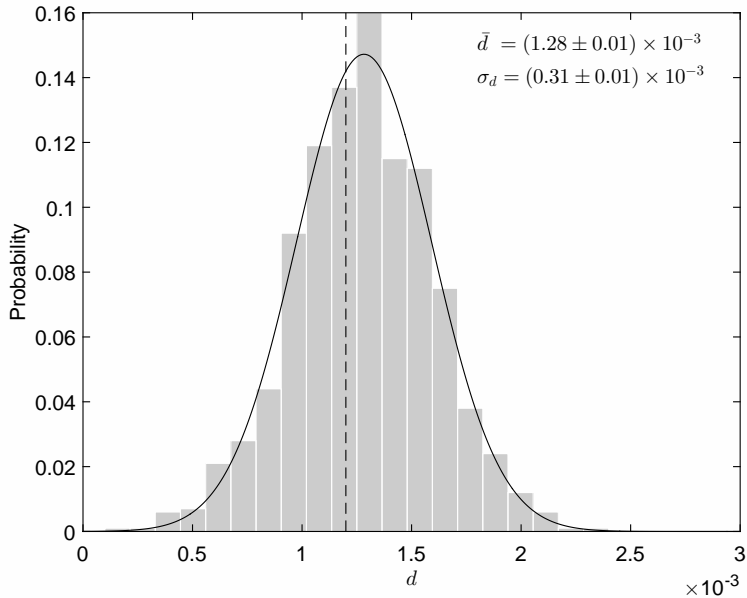}\\
  \caption{The histogram of dipole amplitudes in 1000 simulations with 800 GW events in each simulation. The dashed line is the fiducial dipole amplitude.}\label{fig_aniso_800_d}
\end{figure}

Fig.\ref{fig_aniso_800_d} shows the distribution of dipole amplitudes in 1000 simulations when $N=800$.
The distribution can be fitted by Gaussian function centering at $1.28\times 10^{-3}$, with the standard deviation $\sigma_d=0.31\times 10^{-3}$. This implies that the fiducial dipole amplitude can be correctly reproduced with $\sim 800$ GW events.

\begin{figure}[htbp]
  \centering
  \includegraphics[width=0.65\textwidth]{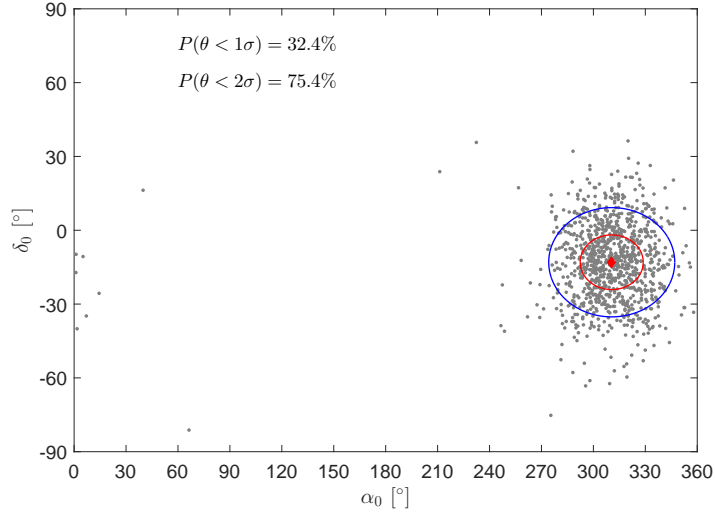}\\
  \caption{The distribution of preferred directions in 1000 simulations with 800 GW events in each simulation. The red and blue error ellipses are the $1\sigma$ and $2\sigma$ confidence regions of Unin2.1, respectively.}\label{fig_aniso_800_lb}
\end{figure}

Fig.\ref{fig_aniso_800_lb} shows the distribution of preferred directions in 1000 simulations when $N=800$.
The red and blue error ellipses are the $1\sigma$ and $2\sigma$ confidence regions of Unin2.1, respectively. From this figures, we can see that the simulated directions are clustered near the fiducial direction. The probabilities of falling into the $1\sigma$ and $2\sigma$ confidence regions of Union2.1 are $32.4\%$ and $75.4\%$ respectively, implying that with about 800 GW events the preferred direction can be correctly recovered.

\begin{figure}[htbp]
  \centering
  \includegraphics[width=0.6\textwidth]{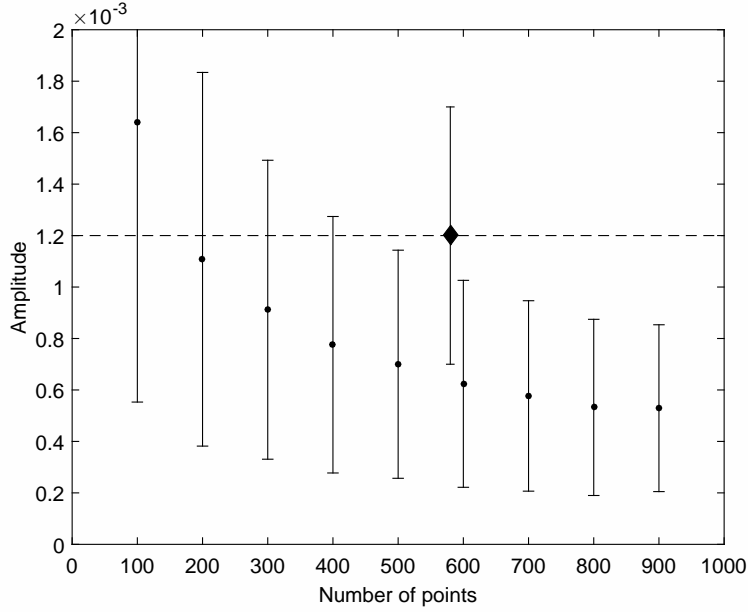}\\
  \caption{The same to Fig.\ref{fig_aniso_all_d} but with the isotropic GW data.}\label{fig_iso_all_d}
\end{figure}

\begin{figure}[htbp]
  \centering
  \includegraphics[width=0.6\textwidth]{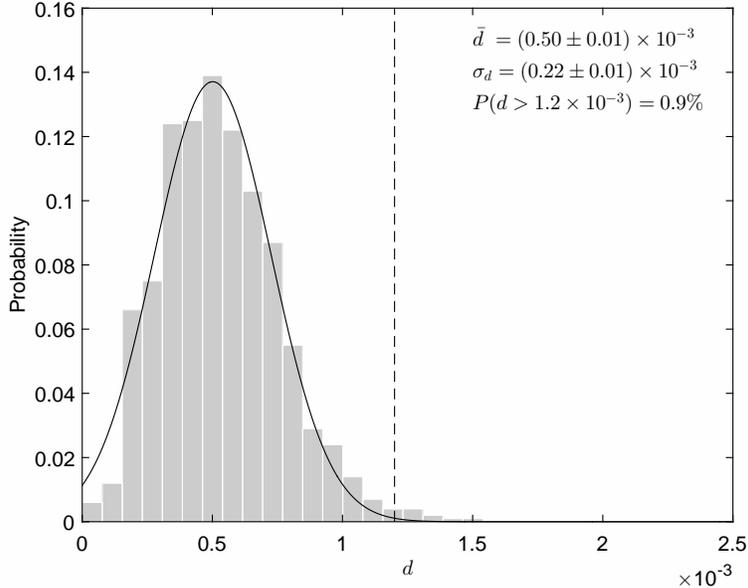}\\
  \caption{The same to Fig.\ref{fig_aniso_800_d} but with the isotropic GW data.}\label{fig_iso_800_d}
\end{figure}

In order to test if the statistical noise can lead to pseudo anisotropic signals, we simulate some isotropic data sets. The simulation procedure is the same to the anisotropic case except that the fiducial dipole amplitude is fixed to zero now. The simulated data is then fitted by the dipole model of equation (\ref{eq:dipole_mu}). As was done above, we simulate different number of GW events, repeat the simulation 1000 times, and calculate the mean value of dipole amplitude. We expect that the fitted dipole amplitude is consistent with zero. The results are shown in Figs.\ref{fig_iso_all_d} and \ref{fig_iso_800_d}.

In Fig.\ref{fig_iso_all_d}, we plot the average dipole amplitude and $1\sigma$ uncertainty as the function of the number of GW events $N$, together with the dipole amplitude and its uncertainty of Union2.1 for comparison. From the figure we may see that the dipole amplitude decreases as the number of GW events increases, but it is not zero within $1\sigma$ uncertainty even if the number of GW events increases to 900. This means that the noise may lead to pseudo anisotropic signal. However, the amplitude of pseudo anisotropic signal is always smaller than the dipole amplitude of Union2.1 if $N>200$. With $\sim 800$ GW events, the former is smaller than the latter at about $2\sigma$ confidence level.

In Fig.\ref{fig_iso_800_d}, we plot the distribution of dipole amplitudes in 1000 simulations in $N=800$ case. The distribution is well fitted by Gaussian function, with the average value $\bar{d}=0.50\times 10^{-3}$ and standard deviation $\sigma_d=0.22\times 10^{-3}$. The probability of pseudo dipole amplitude being larger than the dipole of Union2.1 is only $0.9$ percent. Therefore, if the universe really has an anisotropy with amplitude larger $1\times 10^{-3}$, this anisotropic signal can be tested by $\sim 800$ GW events.

\section{Discussions and conclusions}\label{sec:conclusions}

In this paper, we have investigated the constraint ability of GW events on the anisotropy of the universe using the simulated data from advanced LIGO and Virgo. It is found that the GW data can tightly constrain the anisotropy amplitudes of the universe with $\gtrsim 400$ events if advanced LIGO and Virgo reach to the designed sensitivity. To tightly constrain the preferred direction, however, $\gtrsim 800$ events are needed. The simulated GW events have average uncertainty $0.4$ mag on distance modulus, which is about two factors larger than the Union2.1 compilation of type-Ia supernovae. Here we only considered the GW signals in the inspiral epoch. The GW frequency in the inspiral epoch is about tens Hz, which is bellow the most sensitive frequency of the detectors. The uncertainty can be reduced by consider the GW signals in the merger and ringdown epochs. Here we have assumed that the GW source can be precisely localized and the orbital plane of inspiral is nearly face on. Otherwise the distance of GW source may be correlated with other parameters and the accuracy on the determination of distance gets worse. Therefore, in practise more GW events may be needed in order to achieve the accuracy of Union 2.1. As the improvement of sensitivity, advanced LIGO is expected to detect hundreds of GW events produced by the coalescence of binary black hole systems in the next few years. Therefore, GWs provide a promising complementary tool to supernovae in testing the anisotropy of the universe.

In this paper, we only considered the binary black hole systems as the sources of GWs, while the binary neutron star and binary of neutron star - black hole systems are not considered. This is because the binary of neutron star - black hole systems have not been observed yet, and with the designed sensitivity advanced LIGO and Virgo can only observe the binary neutron star systems at very low redshift. The biggest challenge of using GWs as the standard sirens comes from the localization of GW source. With two detectors, the source can only be localized on a strip of the sky. Even if with three or more detectors, the localization accuracy is at the order of several degrees with present sensitivity. Such an accuracy is far from accurate enough to identify the host galaxy, thus hampers the measurement of redshift. If the GW has electromagnetic counterparts such as short gamma-ray bursts, then the host galaxy can be identified and the redshift can be determined accurately by the follow-up observations. Unfortunately, the merge of binary black hole is expected to have no electromagnetic counterparts. Chen \cite{Chen:2017rfc} pointed out that the redshift can be obtained statistically by analyzing over all potential host galaxies within the localization volume. The redshift inferred in this way, however, adds additional uncertainty to the constraints. This disadvantage of the measurement of redshift can be improved by the on-going third generation detectors with higher sensitivity such as the Einstein Telescope \cite{Einstein Telescope}. However, at present, our researches provide a possible approach to test the standard cosmological model by the advanced LIGO and Virgo detectors in the next years.

\begin{acknowledgments}
This work has been supported by the National Natural Science Fund of China under grant Nos. 11603005 and 11775038, and the Fundamental Research Funds for the
Central Universities project Nos. 106112017CDJXFLX0014 and 106112016CDJXY300002.
\end{acknowledgments}

\end{document}